\documentclass[12pt]{article}

\begin{document}

\title{On the Classification Scheme for Phenomenological Universalities in Growth 
Problems in Physics and Other Sciences}

\author{Marcin Molski\\
Department of Theoretical Chemistry, Faculty of Chemistry\\
A. Mickiewicz University of Pozna\'n,\\
ul. Grunwladzka 6, PL 60-780 Pozna\'n, Poland,\\
email: marcin@rovib.amu.edu.pl}
\maketitle

\begin{abstract}

Comment on "Classification Scheme for Phenomenological Universalities in Growth 
Problems in Physics and Other Sciences" by P. Castorina, P. P. Delsanto and C. Guiot, 
Phys. Rev. Lett. {\bf 96}, 188701 (2006) is presented. It has been proved that the 
West-like function of growth derived by the authors is incorrect and the approach 
does not take into account the growth of the biological systems undergoing atrophy 
or demographic and economic systems undergoing involution or regression. A simple 
extension of the model, which permits derivation of the so far unknown involuted 
Gompertz function of growth is proposed.    

\end{abstract}

The concept of {\it phenomenological universalities} introduced recently by 
Castorina, Delsanto and Guiot (CDG) \cite{Castorina} is a useful tool for investigation of
the nonlinear processes in complex systems whose dynamics is governed by the system of 
equations 
\begin{equation}
\frac{dy(\tau)}{d\tau}=x(\tau)y(\tau)\hskip1.5cm \frac{dx(\tau)}{d\tau}=-\Phi(x).
\label{Eq1}
\end{equation}
Here, $\tau=x(0)t$ denotes dimensionless temporal variable, whereas $\Phi(x)$ is a generating 
function, which extended into a series of x-variable generates different functions of growth 
for a variety of patterns emerging in complex systems in physics, biology and beyond.
For example for $\Phi=x$ one gets the Gompertz function, whereas for $\Phi=x+bx^2$ 
the allometric West-like function is derived 
\cite{Castorina}
\begin{equation}
y(\tau)_{G}=\exp\left[1-\exp(-\tau)\right],\hskip1cm y(\tau)_{W}=\left[1+b-b\exp(-\tau)\right]^{1/(1-b)}.
\label{Eq2}
\end{equation}
Unfortunately, the West-like function obtained by CDG \cite{Castorina} is incorrect as it is 
not a solution of the differential equation (9) in \cite{Castorina}. Additionaly in the limit 
$\lim_{b\to 0}y(\tau)_{W}=1$ it does not produce the Gompertz function as indicated 
by the authors \cite{Castorina}. To explain those inconsistencies Eqs. (\ref{Eq1}) were solved 
for $\Phi=x+bx^2$ employing Maple vs. 7.0 processor for symbolic calculations. The calculations 
provided the correct solution $y(\tau)_{W}=\left[1+b-b\exp(-\tau)\right]^{1/b}$
with power $1/b$ and not $1/(1-b)$ as derived by CDG \cite{Castorina}. 
Employing the Maple one may also prove that 
\begin{equation}
\lim_{b\to 0} \left[1+b-b\exp(-\tau)\right]^{1/b}=\exp\left[1-\exp(-\tau)\right] 
\label{Eq3}
\end{equation}
as it should be.

Although CDG claimed \cite{Castorina} that they {\it .....have developed a simple scheme that allows the classification 
of all the growth problems.....} described by  Eqs. (\ref{Eq1})  this approach does not take 
into account the growth of the biological systems undergoing atrophy or demographic and economic 
systems undergoing involution or regression. In biological systems such a situation appears 
in avian primary lymphoid organs: thymus and bursa of Fabricius as well as in thymus of mammalians.   
To extend the CDG approach and derive an involuted function of growth, we employ 
a linear expansion of the generating function $\Phi(x)=c_1x+c_0$, which includes a constant term $c_0$ 
omitted in the CDG scheme and $c_1$ coefficient, which in the 
previous approach was constrained $c_1=1$ . Additionally, we assume that  
$x(\tau=0)=(1-c_0)/c_1$, which for $c_0=0$ and $c_1=1$ gives the CDG condition $x(0)=1$.
Employing the relationships (\ref{Eq1}) one gets the involuted Gompertz function ($c_0,c_1>0$)
\begin{equation}
y(\tau)=\exp\left\{\frac{1}{c_1^2}\left[1-\exp(-c_1\tau)\right]\right\}\exp\left(-\frac{c_0}{c_1}\tau\right),
\label{Eq11}
\end{equation}
which can be specified in the form applicable to direct fitting of the experimental data
\begin{equation}
y(t)=y_0\exp\left\{\frac{b}{a}\left[1-\exp(-at)\right]\right\}\exp\left(-bct\right).
\label{Eq12}
\end{equation}
Here, we use the correspondences $c_1^2=a/b$, $c_1\tau=at$ and $c=c_0$.
To prove that function (\ref{Eq12}) correctly describes the evolution and involution of organs
undergoing atrophy, we employed it to fit 30 mean absolute weights of the thymus of male
Wistar rats evaluated in the period of 1-780 days. In the calculations we used thymuses sampled
from strain rats of the mean age of 24 months, originating from randomly mated culture (Department
of Toxicology, University of Medical Science in Pozna\'n, Poland). The mean values of the absolute 
thymus weight were fitted to the 4-parameteric function (\ref{Eq12}) using the weighted nonlinear 
least-square routine with statistical weights taken as inverse squares of uncertainties $u(i)$ of the 
thymus weights. As the criteria of goodness of the fit,  the standard deviation $\sigma$ 
and normalized standard deviation $\hat{\sigma}$ were employed. The calculations provided the 
values of the parameters $y_0=0.0105(16)$ [g], $a=0.0472(44)$ [1/day], $b=0.1914(194)$ [1/day]  
and $c=0.0140(38)$ [1/day]  reproducing the experimental data with $\sigma=0.011$ [g] 
and $\hat{\sigma}=1.64$. All parameters fitted are statistically well evaluated and the correlation
coefficients between them cover a satisfactory range $(0.6455-0.8971)$. 

The results obtained indicate that the $U_1$ class introduced by CDG \cite{Castorina} should be 
divided into two subclasses $U_{11}$ - representing the sigmoidal Gompertzian growth, and $U_{12}$ - 
including the involuted Gompertzian growth. The $U_{12}$ solution (\ref{Eq12}) is a very important 
contribution to the classes obtained previously as it is usefull to fit the data for the systems 
undergoing atrophy. 
The sigmoidal (S-shaped) functions (Gompertz, West, logistic, Richards etc.)  cannot be apply
to this aim as they correctly describe the ideal situation in which exponential growth is exponentially 
retarded and saturated as time continues.

\section*{Acknowledgments}

The author wishes to thank Prof. Renata Brelinska from the Department of Histology and Embryology 
at University of Medical Sciences in Pozna\'n, Poland for providing the experimental data employed in 
calculations. The experimental part of this comment was supported by grant KBN 2 PO5A 022 28.

\end{document}